\documentclass{article}
\usepackage{spconf,amsmath,graphicx,hyperref}
\usepackage{multirow}
\usepackage{booktabs}
\usepackage{adjustbox}
\usepackage{pifont}
\usepackage{amssymb}
\usepackage{tipa}
\usepackage{CJKutf8}
\usepackage{makecell}

\let\OLDthebibliography\thebibliography
\renewcommand\thebibliography[1]{
  \OLDthebibliography{#1}
  \setlength{\parskip}{2pt}
  \setlength{\itemsep}{2pt plus 0.4ex}
}

\title{A Bottom-up Framework with Language-universal Speech Attribute Modeling for Syllable-based ASR}
\name{%
\begin{tabular}{@{}c@{}}
Hao Yen$^{1}$, Pin-Jui Ku$^{1}$, Sabato Marco Siniscalchi$^{1,2,3}$, Chin-Hui Lee$^{1}$
\end{tabular}}
\address{$^1$Georgia Institute of Technology, USA\\$^2$Università degli Studi di Palermo, Italy\\$^3$Norwegian University of Science and Technology, Norway}
\begin{document}
\ninept
\maketitle
\begin{abstract}
We propose a bottom-up framework for automatic speech recognition (ASR) in syllable-based languages by unifying language-universal articulatory attribute modeling with syllable-level prediction. The system first recognizes sequences or lattices of articulatory attributes that serve as a language-universal, interpretable representation of pronunciation, and then transforms them into syllables through a structured knowledge integration process. We introduce two evaluation metrics, namely Pronunciation Error Rate (PrER) and Syllable Homonym Error Rate (SHER), to evaluate the model's ability to capture pronunciation and handle syllable ambiguities. Experimental results on the AISHELL-1 Mandarin corpus demonstrate that the proposed bottom-up framework achieves competitive performance and exhibits better robustness under low-resource conditions compared to the direct syllable prediction model. Furthermore, we investigate the zero-shot cross-lingual transferability on Japanese and demonstrate significant improvements over character- and phoneme-based baselines by 40\% error rate reduction.
\end{abstract}

\vspace{-.3cm}
\section{Introduction}
Traditional top-down ASR systems \cite{Jelinek1976} follow a model-driven approach, in which the acoustic signal is decoded using statistical acoustic models, such as HMMs or DNNs \cite{Povey2011,Lee1988}, a pronunciation lexicon, and a language model to obtain the most likely word sequence. These systems rely on higher-level constraints and predefined word structures, making them highly effective in large-vocabulary read speech recognition tasks but less robust in spontaneous speech or low-resource scenarios. In contrast, the bottom-up ASR approach \cite{Lee2013} is motivated by human speech recognition (HSR) \cite{Allen1995,Lippmann1997} theories, which suggest that the linguistic identity of speech sounds is determined through the incremental accumulation of acoustic evidence. Bottom-up ASR frameworks reverse the hierarchy in top-down ASR systems by focusing first on the detection of fundamental speech events, such as articulatory attributes \cite{Chomsky1968,Clements1985}, and then composing higher-level linguistic structures from these observations.

Bottom-up ASR architectures offer several compelling advantages compared to traditional approaches, including top-down and end-to-end (E2E) ASR systems \cite{Xiao2018,Watanabe2018,Watnabe2017hybrid,Radford2022}. First, they are highly modular and interpretable. Because processing is organized into distinct detectors and integration stages, the system provides rich diagnostic information at each level. It becomes easier to analyze errors, where one can inspect which attribute detector failed or which combination of features led to a wrong predictions. This advantage is largely impossible in an top-down or E2E recognizer. This modularity also means one can evaluate and tune individual components, e.g., the articulatory attribute acoustic models, in isolation using feature-specific test sets. Second, designing and refining an individual component is often more straightforward than modifying a large and black-box ASR system. Researchers can focus on one module at a time and incorporate domain knowledge to make it as reliable as possible. Over time, this leads to a library of expert classifiers that collectively cover the acoustic space. Third, bottom-up systems naturally exploit multi-level evidence and can be more robust in challenging conditions \cite{Kirchhoff1998}. Since attributes are modeled independently across the speech signal, the framework can capture overlapping and redundant cues. Fourth, bottom-up ASR is advantageous in low-resource settings and for unseen inputs. Since it does not rely strictly on a fixed vocabulary or massive language model training, it can recognize patterns not seen in training by piecing together detected pronunciation units \cite{Hon1992,Yen2025}. A number of studies have demonstrated the effectiveness of bottom-up frameworks. For instance, Bromberg et al. \cite{Bromberg2007} trained neural detectors for 14 phonological features and used a CRF-based integration model, achieving competitive results on TIMIT. Metze and Waibel \cite{Metze2002AFS} showed that flexible streaming architectures leveraging articulatory features can improve robustness in reverberant conditions. Liu’s landmark-based approach \cite{Liu1996} and Juneja’s event-based recognition system \cite{Juneja2002} further demonstrate the feasibility of using detected speech events to guide recognition. These validate the viability of bottom-up systems as an alternative or complement to traditional top-down ASR.

In this work, we extend previous universal attribute modeling efforts, and present a framework toward building a bottom-up ASR. The design concept of the proposed system consists of a comprehensive inventory of fundamental attribute units, a pre-trained self-supervised acoustic model, and a knowledge integration process. The acoustic model initially generates pronunciation information (attribute posteriors), followed by a knowledge-driven process to select the appropriate knowledge source. Our experiments on Mandarin Chinese demonstrate that the bottom-up framework can capture pronunciation more accurately than traditional E2E models, and achieve superior performance once training data is limited. Furthermore, cross-lingual experiment to Japanese showcases the inherent language-universal property of the proposed framework with improved syllable recognition accuracy of over 40\% compared to commonly used units like characters and phonemes.

\vspace{-.3cm}
\section{Related Work}
\subsection{Syllable-based Languages and Speech Recognition}
In languages, such as English, character or subword units are commonly adopted as modeling targets for speech recognition, as these units generally align well with phonological structure. However, in many other languages, particularly those with logographic or syllabic writing systems, characters function primarily as semantic or orthographic symbols and lack a direct correspondence to pronunciation. Consequently, character-based modeling in these languages fails to incorporate phonological information into the acoustic model. Phoneme-based approaches also encounter limitations, particularly in languages with strong supra-segmental features, where they struggle to capture tonal and prosodic variations critical for accurate recognition. Given these limitations, syllables offer a linguistically coherent and acoustically meaningful alternative for speech recognition in such languages, often referred to as syllable-based languages. Both Mandarin Chinese and Japanese belong to this category. This strong alignment between orthography and pronunciation makes syllables a compelling choice for acoustic modeling in these languages. In most cases, better syllable recognition leads to more accurate recognition of orthographic units.


Syllable-based acoustic modeling has attracted increasing interest in recent years. Ganapathiraju et al. \cite{Ganapathiraju2001} first demonstrated that syllable-based acoustic models could outperform traditional phone-based models in large-vocabulary tasks using HMM/GMM system. In the context of Mandarin Chinese, many prior studies \cite{Wu2007,Qu2017,Zhou2018ACO,Zhou2018,Yuan2021} demonstrate that syllable-level models can outperform context-independent phoneme models and even approach or surpass state-of-the-art context-dependent baselines. In Japanese ASR, traditional systems historically used mora units \cite{Kubozono1989,Tomokiyo1997,Takahashi2002,Hiroshi2014} for acoustic modeling. It is important to note that our work differs from \cite{Zhang2011} in that we focus on syllables rather than phonemes, as syllables exhibit a closer correspondence to orthographic units in written text. Moreover, we extend our investigation beyond Mandarin to broader syllable-based languages, including Japanese, to highlight the language-universal nature of the proposed framework.


\begin{figure}
    \centering
    \includegraphics[width=\columnwidth]{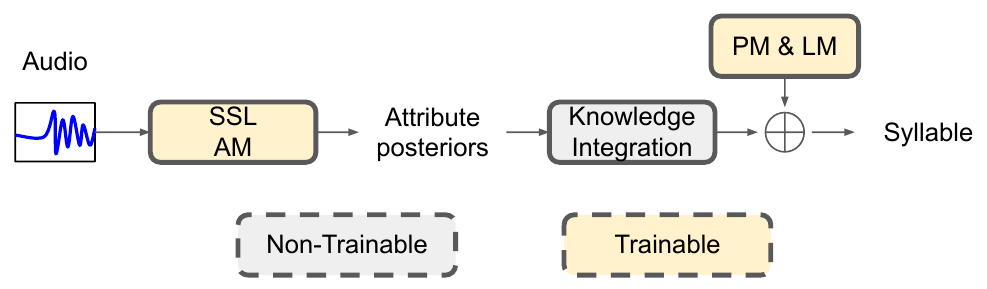}
    \caption{Overview of the bottom-up framework for syllable-based ASR. "AM" refers to acoustic model, "PM" denotes pronunciation model, and "LM" stands for language model.}
    \label{fig:csr_framework}

\end{figure}
\section{Proposed Bottom-up Framework for Syllable-based Speech Recognition}

\subsection{Attribute Recognition and Knowledge Integration}
The proposed bottom-up framework is illustrated in Figure \ref{fig:csr_framework}. The system begins with an acoustic model (AM) built on WavLM \cite{Chen2022wavlm} to predict speech attribute posteriors directly from audio. These posteriors represent articulatory knowledge sources and serve as language-universal representation. Following this, a knowledge integration step is applied to select and combine the appropriate knowledge sources as well as level of attribute granularity. This step enables flexible modeling by empirically choosing which subset of attributes to include based on the desired trade-off between representational compactness and discriminative power. The attribute sequences are then transformed into syllables, serving as the bridge to higher-level linguistic units. The final recognition output is produced via decoding with a pronunciation model (PM) and refined with an optional language model (LM) trained on syllable sequences, which can be viewed as auxiliary knowledge sources.

\subsection{Universal Attribute Inventory}
\label{sec:our_attr_model}
The set of speech attributes used in this work are distinctive features \cite{Chomsky1968,Clements1985,Fant1973,Ladefoged2012}. The attribute inventory normally consists of the following categories: \textit{manner of articulation} (M), \textit{place of articulation} (P), \textit{voicing} (V), and \textit{aspiration}. These attributes are designed to capture articulatory properties that are consistent across languages, providing a compact yet expressive representation of speech sounds. The \textit{manner of articulation} includes eleven features and the \textit{place of articulation} set comprises ten features. In addition, \textit{voicing} consists of voiced or voiceless, and \textit{aspiration} consists of aspirated or unaspirated. As pointed out in \cite{Tang2003}, vowels and consonants cannot be mapped into a common linguistic space, because place of articulation has been differently defined for them. Therefore, we introduce two articulatory dimensions: Height (H) and Backness (B). Height captures the vertical tongue position during vowel articulation and consists of seven classes: high, semi-high, upper-mid, mid, lower-mid, semi-low, and low. Backness, on the other hand, represents the horizontal tongue position and is categorized as front, central, or back. The inventory of all universal attribute units adopted in this paper is shown in Table \ref{tab:attributes}.

\begin{table}[t!]
    \centering
    \caption{Full inventory of speech attributes used in this paper.}
    \label{tab:attributes}
    \adjustbox{width=\columnwidth}{
    \begin{tabular}{c|l}
        \toprule
        \textbf{Category} & \textbf{Attributes} \\ \midrule \midrule
         Manner (M) &  \makecell[tl]{nasal, stop, affricate, fricative, flap, trill, approximant, \\ click, ejective, implosive, vowel} \\ \midrule
         Place (P) & \makecell[tl]{bilabial, labiodental, dental, alveolar, palato-alveolar, \\
         retroflex, alveolo-palatal, palatal, velar, uvular, glottal, vowel} \\ \midrule
         Voicing (V) & voiced, voiceless \\ \midrule
         Aspiration (A) & aspirated, unaspirated \\ \midrule \midrule
         Height (H) & \makecell[tl]{high, semi-high, upper-mid, mid, lower-mid, semi-mid, low} \\ \midrule
         Backness (B) & front, central, back \\ \bottomrule
    \end{tabular}
    }
\end{table}


\subsection{Mapping Syllables to Attribute Sequences}
\label{sec:syl2attr}
In Mandarin Chinese, each syllable consists of an initial consonant, or null initial, and a final containing the vowel or vowel combination, plus any semivowel glide or syllable-final nasal. We convert each syllable into an articulatory attribute sequence by mapping the initial and final separately, then concatenating their attributes. Initials are described by their manner and place of articulation, with an explicit aspiration feature to distinguish aspirated from unaspirated stops and affricates, which is essential for contrasts such as /b/ vs. /p/, /d/ vs. /t/, and /zh/ vs. /ch/, since Mandarin uses aspiration rather than voicing as the primary contrast. Vowels use the height and backness features. Standard Japanese syllables share the same CV structure, with consonants categorized by manner and place of articulation, and contrasting primarily by voicing (V) \cite{Akamatsu1997}, allowing pairs such as /k/ vs. /g/, /t/ vs. /d/, and /s/ vs. /z/ to be distinguished in attributes \cite{Gao2018}. Vowels use the same height and backness features as in Mandarin. Japanese also has a moraic nasal that may appear syllable-final, realized as [m], [n], or [\textipa{N}]; unlike \cite{Kubozono1989,Ogata2003}, we merge it with the preceding syllable, assigning it a nasal manner and an appropriate place attribute. 


\section{Experiments and Results}
\subsection{Datasets \& Experimental Settings}
The AISHELL-1 corpus \cite{Bo2017} is a Mandarin Chinese speech dataset containing approximately 178 hours of transcribed speech recorded from 400 speakers using high-fidelity microphones. For our experiments, we use the official 150-hour training set and an 18-hour development set for validation and early stopping. Evaluation is performed on the 10-hour test set, which includes 7,176 sentences. In the experiments, we will focus on toneless syllable recognition which has a total of 408 units. To simulate low-resource training conditions, we create three training subsets by randomly selecting 25\%, 10\%, and 5\% of the original full training set. For Japanese, we use the JSUT corpus \cite{Sonobe2017} for evaluation. This corpus provides Japanese text transcriptions paired with read-style speech recordings performed by a native Japanese female speaker. The full corpus contains approximately 10 hours of speech covering various speaking scenarios. In our experiments, we use the “basic5000” subset, which includes 5,000 sentences designed to cover all daily-use Japanese characters. We follow the official ESPnet \cite{Watanabe2018} JSUT recipe and use the first 500 sentences in the provided list as the test set. 

We adopt a publicly available pre-trained self-supervised model as our acoustic model across all experiments, namely WavLM \cite{Chen2022wavlm}. We use the "large" version, which comprises approximately 311 million parameters, featuring 24 Transformer \cite{Vaswani2017} encoder layers with an embedding dimension of 1024 and 16 attention heads. The model is pre-trained on 94k hours of unannotated speech data from multiple sources. All models are trained using the AdamW optimizer \cite{Loshchilov2018} with $\beta_1=0.9$ and $\beta_2=0.98$ and optimized with the CTC loss function. To support decoding with CTC model, we trained separate KenLM language models tailored to each modeling unit.


\vspace{-.1cm}
\subsection{Pronunciation and Syllable Homonym Error Rate}
To evaluate how accurately the model captures the underlying pronunciation, independent of orthographic symbols, we introduce a metric called Pronunciation Error Rate (PrER). This metric quantifies the recognition quality at the level of articulatory attributes, offering a more direct evaluation of the acoustic modeling capability. Since not all model outputs are in the form of attribute sequences, with higher-level tokens such as syllables, we need to first convert all predicted outputs into attribute sequences using the mapping procedure described in Section \ref{sec:syl2attr}. PrER enables a uniform, articulatory-level comparison of pronunciation quality across different modeling units, and a lower value indicates that the model has more accurately captured the phonetic structure of the spoken input. In practice, PrER can be calculated on different acoustic resolution, including a single category of attribute or any combination of attributes.

At the syllable level, in addition to the conventional Syllable Error Rate (SER), we introduce the Syllable Homonym Error Rate (SHER) to evaluate recognition performance under the inherent homonymy when transforming attribute posteriors to syllable sequence. In many cases, a single attribute sequence can correspond to multiple syllables due to overlapping articulatory representations, especially when the attribute inventory lacks full discriminative power. To account for this, SHER treats all syllables sharing the same underlying attribute sequence as equivalent, effectively providing a relaxed version of SER. This metric serves as a lower bound on the achievable SER of any attribute-based system, representing the ideal case in which homonymous syllables are perfectly disambiguated in a downstream rescoring or mapping stage.


\begin{table}[t]
\vspace{-.3cm}
\centering
\caption{PrER(\%) on test set of AISHEll-1 for different categories of speech attributes. No language model is applied during decoding.}
\label{tab:aishell_prer}
\adjustbox{width=0.8\columnwidth}{
\begin{tabular}{l|ccccc}
\toprule
\multirow{2}{*}{\textbf{System}} & \multicolumn{5}{c}{\textbf{PrER (\%)}} \\
& M & P & H & B & A\\
\midrule \midrule
$\text{WavLM}_\text{syl}$ & 0.78 & 0.78 & 0.79 & 0.76 & 0.53 \\
$\text{WavLM}_\text{attr}$ (ours)   &  \textbf{0.72} &  \textbf{0.75} &  \textbf{0.77} &  \textbf{0.65} &  \textbf{0.51}\\
\bottomrule
\end{tabular}
}
\vspace{-.3cm}
\end{table}

\begin{table}[t]
\centering
\caption{Experimental results on the AISHELL-1 test set, reporting SER and SHER. For the proposed bottom-up system, different knowledge sources (attribute categories) are added incrementally.}
\label{tab:aishell_all}
\adjustbox{width=0.9\columnwidth}{
\begin{tabular}{l|l|cc}
\toprule
\multicolumn{1}{c|}{\textbf{System}} & \multicolumn{1}{c|}{\textbf{Know. Source}} & \multicolumn{1}{c}{\textbf{SER(\%)}} & \multicolumn{1}{c}{\textbf{SHER(\%)}} \\\midrule \midrule
$\text{WavLM}_\text{syl}$ & -- & \textbf{2.36}    &  2.33     \\ \midrule \midrule
\multirow{3}{*}{$\text{WavLM}_\text{BU}$ (ours)} & M+P  & 47.62  & --  \\
& M+P+H  & 9.83  & -- \\
& M+P+H+B+A  & 2.47  & \textbf{2.12}  \\ 
\bottomrule
\end{tabular}
}
\end{table}

\subsection{Recognition Results on AISHELL-1}
Table \ref{tab:aishell_prer} and \ref{tab:aishell_all} present the PrER, SER, and SHER results on the AISHELL-1 test set using the full 150-hour training data. Specifically, $\text{WavLM}_\text{syl}$ denotes the baseline model trained to predict syllables directly, while $\text{WavLM}_\text{BU}$ represents our proposed bottom-up framework. For evaluating PrER, no language model is applied during decoding to isolate the pure acoustic modeling performance. As shown in the results, $\text{WavLM}_\text{BU}$ consistently outperforms the $\text{WavLM}_\text{syl}$ in terms of PrER across all categories of speech attributes. These results highlight the advantage of the bottom-up framework in capturing fine-grained phonetic details that conventional syllable-level models may miss, underscoring its potential for more accurate and language-universal pronunciation modeling.

In addition, we can make several observations from Table \ref{tab:aishell_all}. First, the intermediate SER results of $\text{WavLM}_\text{BU}$ illustrate the benefit of progressively adding the articulatory knowledge source. The SER drops sharply from 47.62\% when using only manner and place, to 9.83\% when adding vowel height, and further down to 2.47\% with the complete attribute set. This trend confirms that finer-grained articulatory details effectively boost recognition accuracy. The best bottom-up model using the full knowledge source (M+P+H+B+A) achieves a SER of 2.47\%, which is slightly higher than $\text{WavLM}_\text{syl}$ (2.36\%). To better isolate the true acoustic modeling capability, we also compute SHER based on the complete knowledge source (M+P+H+B+A), meaning two syllables are considered equivalent if their underlying attribute sequences are identical across all categories. Under this relaxed evaluation, $\text{WavLM}_\text{BU}$ achieves a lower SHER of 2.12\%, outperforming $\text{WavLM}_\text{syl}$ with 2.33\%. These results suggest that most of the errors in the bottom-up system occur not from misrecognition of the phonetic content, but from ambiguity in syllable homonym. In contrast, for $\text{WavLM}_\text{syl}$, the close alignment between its SER and SHER indicates that its errors stem directly from acoustic misrecognition, which are inherently more difficult to correct during post-processing or rescoring. 

Table \ref{tab:aishell_exp} presents a qualitative example from the AISHELL-1 test set. In the example, $\text{WavLM}_\text{syl}$ misrecognizes both /xian/ and /jin/ as /qian/ and /jie/, respectively, and fails to capture the syllable /yi/ and the nasal ending of /deng/. In contrast, although $\text{WavLM}_\text{BU}$ outputs /xian/ as /xuan/, the two syllables share the same pronunciation in terms of all M, P, H, B, and A. This contrast underscores that with improved rescoring and additional knowledge sources, our proposed bottom-up approach has better chance of correcting the mistakes.

\begin{table}[t]
\vspace{-.3cm}
    \centering
    \caption{A qualitative example from the AISHELL-1 test set.}
    \label{tab:aishell_exp}
    \adjustbox{width=0.8\columnwidth}{
    \begin{tabular}{c|c}
        \toprule
        \textbf{Model} & \textbf{Syllables} \\ \midrule
        Ground Truth & /... wu \textbf{xian yi jin deng} wei jie kou/ \\
        $\text{WavLM}_\text{syl}$ & /... wu \textbf{qian jie de} wei jie kou/\\
        $\text{WavLM}_\text{BU}$ & /... wu \textbf{xuan} yi jin deng wei jie kou/\\
        \bottomrule
    \end{tabular}
    }
    \vspace{-.3cm}
\end{table}

\subsection{Ablation Study on Low-resource Training}
\begin{figure}[b!]
    \vspace{-.3cm}
    \centering
    \includegraphics[width=0.8\columnwidth]{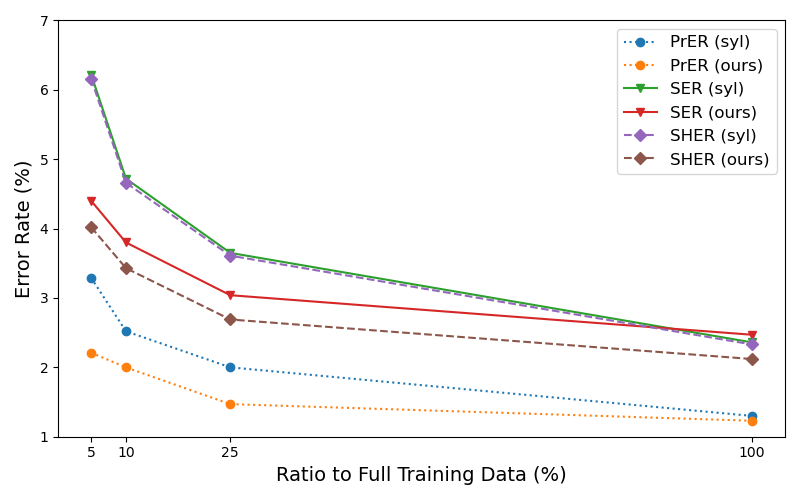}
    \caption{Experimental results (PrER, SER, and SHER) of using different amount of training data. The results for $\text{WavLM}_\text{syl}$ (syl) and $\text{WavLM}_\text{BU}$ (ours) are shown.}
    \label{fig:csr_lr}
\end{figure}

To further evaluate the robustness of the proposed bottom-up framework under limited supervision, we conduct experiments on the AISHELL-1 dataset using only 5\%, 10\%, and 25\%, of the available training data. Figure \ref{fig:csr_lr} shows the comparative results across three evaluation metrics: PrER, SER, and SHER. As shown in the plot, the bottom-up system consistently achieves lower PrER than the syllable-based model across all training sizes. This highlights the model’s superior ability to learn pronunciation-relevant acoustic features, especially under low-resource conditions. Since PrER is measured without any language model, this advantage can be attributed solely to the stronger acoustic modeling of articulatory attributes with the bottom-up system, which enables more data-efficient learning of pronunciation structure.

While the syllable-based model outperforms the bottom-up system in terms of SER when the full training data is used, the bottom-up system actually surpass the syllable-based baseline under low-resource conditions where less than 25\% of the training data is presented. This advantage is likely due to the significantly higher number of syllable units, which makes it difficult to adequately train all units when data is limited, whereas the bottom-up system benefits from a more compact and generalizable unit inventory. In addition, we observe a similar pattern to that in Table \ref{tab:aishell_all}, where the SHERs of $\text{WavLM}_\text{syl}$ closely matches its SERs across all training conditions. In contrast, the SHERs of $\text{WavLM}_\text{BU}$ remain consistently lower than their corresponding SERs in all scenarios. 


\subsection{Zero-shot Cross-lingual Transfer to Japanese}

Table \ref{tab:jsut_prer} shows the PrER results for zero-shot Japanese syllable recognition using our proposed bottom-up framework trained on different source languages. Besides the $\text{WavLM}_\text{BU}$ model trained on AISHELL-1, we also train an additional model on the Multilingual LibriSpeech (MLS) \cite{Pratap2020mls} with eight Indo-European languages, following the same bottom-up manner. We can make several observations from these results. First, the model trained solely on Mandarin Chinese demonstrates a moderate level of cross-lingual transfer capability. In comparison, the model trained on MLS achieves substantially lower PrER on Japanese despite using less total training data since it is exposed to eight languages, giving it a broader, more universal phonetic foundation. The result is a stronger zero-shot performance: even with only 80 hours of training data, the multilingual model’s Japanese PrER is lower, underscoring the benefit of phonetic diversity in the training corpus. Despite never seeing Japanese data during training, it is able to predict Japanese speech attributes to a certain degree. This confirms that a language-universal attribute inventory can generalize beyond the source language.



\begin{table}[t]
\vspace{-.3cm}
\centering
\caption{PrER(\%) on subset of JSUT dataset ("basic5000") for different categories of speech attributes, from different source language ($S$), including 150 hours of Mandarin Chinese (\texttt{ZH}) and 80 hours of Multilingual LibriSpeech (\texttt{MLS}), to target ($T$) Japanese (\texttt{JP}).}
\label{tab:jsut_prer}
\adjustbox{width=0.9\columnwidth}{
\begin{tabular}{l|c|ccccc}
\toprule
\multirow{2}{*}{$S \rightarrow T$} & \textbf{\# Train} & \multicolumn{5}{c}{\textbf{PrER (\%)}} \\
& \textbf{(hrs)} & M & P & V & H & B \\
\midrule \midrule
\texttt{ZH} $\rightarrow$ \texttt{JP} & 150 & 33.76 & 32.85 & 22.55 & 34.71 & 31.22\\
\texttt{MLS} $\rightarrow$ \texttt{JP}  & 80 &  18.16 &  19.87 &  12.34 & 23.39 & 16.50 \\
\bottomrule
\end{tabular}
}

\end{table}

\begin{table}[t]
\vspace{-.2cm}
\centering
\caption{SER(\%) on the subset of JSUT dataset ("basic5000") test set. All the systems are trained on 80 hours MLS dataset and tested on Japanese. For the bottom-up system, different knowledge sources (attribute categories) are added incrementally.}
\label{tab:jsut_ser}
\adjustbox{width=0.7\columnwidth}{
\begin{tabular}{l|l|c}
\toprule
\multicolumn{1}{c|}{\textbf{System}} & \multicolumn{1}{c|}{\textbf{Know. Source}} & \multicolumn{1}{c}{\textbf{SER(\%)}} \\\midrule \midrule
$\text{WavLM}_\text{char}$ & character & 67.79 \\
$\text{WavLM}_\text{phn}$ & phoneme & 64.92 \\ \midrule \midrule
\multirow{3}{*}{$\text{WavLM}_\text{BU}$ (ours)} & M+P  & 79.43   \\
& M+P+H  & 63.36  \\
& M+P+V+H+B  & 40.08 \\ 
\bottomrule
\end{tabular}
}
\vspace{-.3cm}
\end{table}

Table \ref{tab:jsut_ser} shows the SER results of using $\text{WavLM}_{\text{BU}}$ to perform zero-shot transfer to Japanese, comparing to character- and phoneme-based systems. Here all our systems are trained on the 80 hours MLS dataset as it shows better PrER in Table \ref{tab:jsut_prer}. From the results, we observe that although characters and phonemes can be viewed as relatively language-universal units, the systems $\text{WavLM}_{\text{char}}$ and $\text{WavLM}_{\text{phn}}$ struggle with cross-lingual transfer, resulting in high SERs of 67.79\% and 64.92\%, respectively. In contrast, $\text{WavLM}_{\text{BU}}$ demonstrates similar patterns as shown in Mandarin Chinese with consistent improvement as more articulatory knowledge sources are incorporated. Specifically, with the full categories of attributes (M+P+V+H+B), the SER drops to 40.08\%. This demonstrates the potential for cross-lingual transfer to unseen languages and highlights the effectiveness of leveraging language-universal speech attributes combined with minimal language-specific knowledge sources.

\vspace{-.2cm}
\section{Conclusion}
\vspace{-.1cm}
We have presented a bottom-up framework for ASR in syllable-based languages by leveraging a language-universal set of articulatory attributes as the fundamental modeling unit. Through detailed analysis on Mandarin Chinese and Japanese, we demonstrated that the proposed approach effectively models fine-grained pronunciation, achieves competitive syllable recognition performance, and maintains robustness under limited supervision. Furthermore, our cross-lingual transfer experiments confirmed the framework’s inherent ability to generalize to unseen syllable-based languages. These findings underscore the potential of bottom-up attribute modeling as a versatile and scalable foundation for multilingual ASR systems, particularly in low-resource or zero-shot settings.


\newpage
\footnotesize
\bibliographystyle{IEEEbib}
\bibliography{strings,refs}

\end{document}